\documentclass[%
reprint,
amsmath,amssymb,
aps,
]{revtex4-1}

\usepackage{color}

\usepackage{hyperref}
\usepackage{graphicx}
\usepackage{dcolumn}
\usepackage{bm}

\begin{document}
	
	\preprint{APS/123-QED}
	
	\title{Multi-scale brain networks}
	
	\author{Richard F. Betzel$^1$}
	\author{Danielle S. Bassett$^{1,2}$}
	\email{dsb @ seas.upenn.edu}
	\affiliation{
		$^1$Department of Bioengineering, University of Pennsylvania, Philadelphia, PA, 19104
	}
	\affiliation{
		$^2$Department of Electrical and Systems Engineering, University of Pennsylvania, Philadelphia, PA, 19104
	}
	
	\date{\today}
	\begin{abstract}
		The network architecture of the human brain has become a feature of increasing interest to the neuroscientific community, largely because of its potential to illuminate human cognition, its variation over development and aging, and its alteration in disease or injury. Traditional tools and approaches to study this architecture have largely focused on single scales -- of topology, time, and space. Expanding beyond this narrow view, we focus this review on pertinent questions and novel methodological advances for the multi-scale brain. We separate our exposition into content related to multi-scale topological structure, multi-scale temporal structure, and multi-scale spatial structure. In each case, we recount empirical evidence for such structures, survey network-based methodological approaches to reveal these structures, and outline current frontiers and open questions. Although predominantly peppered with examples from human neuroimaging, we hope that this account will offer an accessible guide to any neuroscientist aiming to measure, characterize, and understand the full richness of the brain's multiscale network structure -- irrespective of species, imaging modality, or spatial resolution.
	\end{abstract}
	
	\maketitle
	\section{Introduction}
	
	Over the past decade, the neuroimaging community has witnessed a paradigm shift. The view that localized populations of neurons and individual brain regions support cognition and behavior has gradually given way to the realization that connectivity matters \cite{bassett2006small,sporns2011networks,bullmore2011brain,bressler2010large,park2013structural}. The complex spatiotemporal activity patterns that have been associated with cognition are underpinned by expansive networks of anatomical connections \cite{hagmann2008mapping,hermundstad2013structural,goni2014resting}. This shift has occurred in parallel with the maturation of another field, network science, which has made available a large set of analytic tools and frameworks for characterizing the organization of complex networks \cite{newman2003structure,borgatti2009network,barabasi2016network}.
	
\begin{figure*}[t]
		\begin{center}
			\centerline{\includegraphics[width=0.9\textwidth]{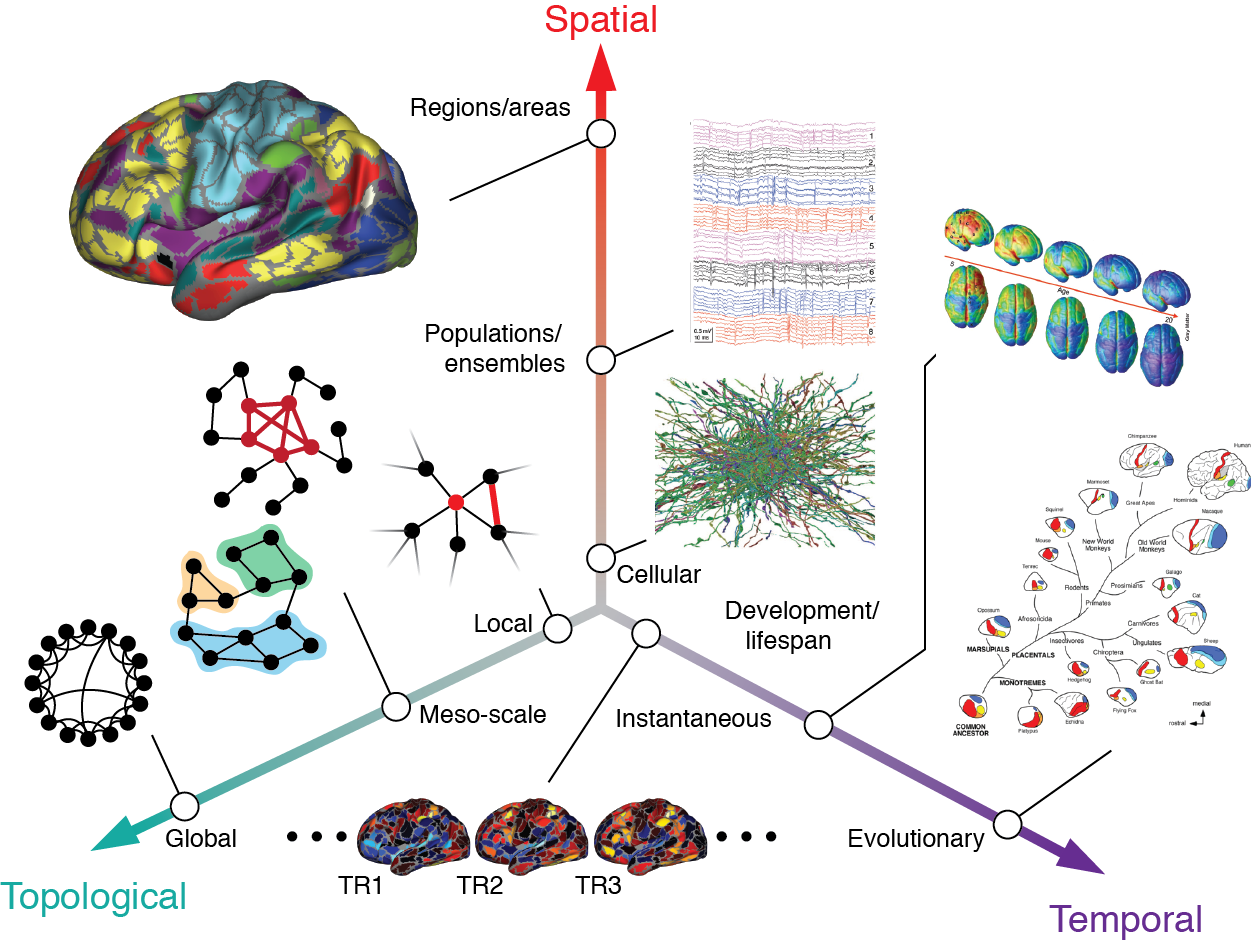}}
			\caption{\textbf{The multi-scale brain.} Brain networks are organized across multiple spatiotemporal scales and also can be analyzed at topological (network) scales ranging from individual nodes to the network as a whole. Images of neuronal ensemble recordings, segmented axons, brain evolution, and gray-matter development adapted with permission from \cite{buzsaki2004large,beyer2013exploring,krubitzer2009search,gogtay2004dynamic}.}
			\label{figure:overviewFigure}
		\end{center}
	\end{figure*}

	As with any new field, the best practices for constructing and analyzing brain networks are still evolving. Among recent developments is the understanding that brain networks are fundamentally multi-scale entities \cite{bassett2013multiscale}. The meaning of ``scale'' can vary depending on context; here we focus on three possible definitions relevant to the study of brain networks. First, a network's \emph{spatial scale} refers to the granularity at which its nodes and edges are defined and can range from that of individual cells and synapses \cite{jarrell2012connectome, shimono2015functional, schroeter2015emergence, lee2016anatomy} to brain regions and large-scale fiber tracts \cite{bullmore2011brain}. Second, networks can be characterized over \emph{temporal scales} with precision ranging from sub-millisecond \cite{khambhati2015dynamic, burns2014network} to that of the entire lifespan \cite{zuo2010growing, betzel2014changes, gu2015emergence}, to evolutionary changes across different species \cite{van2016comparative}. Finally, networks can be analyzed at different \emph{topological scales} ranging from individual nodes to the network as a whole \cite{stam2007graph,bullmore2009complex,rubinov2010complex}. Collectively, these scales define the axes of a three-dimensional space in which any analysis of brain network data lives (Fig.~\ref{figure:overviewFigure}). Most brain network analyses exist as points in this space -- i.e. they focus on networks defined singularly at one spatial, temporal, and topological scale. We argue that, while such studies have proven illuminating, in order to better understand the brain's true multi-scale, multi-modal nature, it is essential that our network analyses begin to form bridges that link different scales to one another.
	
	In this review, we focus on two specific aspects of the multi-scale brain. First, we present and discuss variations of network algorithms (particularly, \emph{community detection}) that make it possible to describe a network at multiple topological scales \cite{porter2009communities,fortunato2010community}. We choose to focus on community detection -- which we define carefully in the next section -- because it encompasses one of the most frequently used set of tools capable of extracting and characterizing network organization across a continuous range of scales. We do, of course, make mention of other alternatives. Next, we discuss the topic of multi-scale temporal networks and a set of multi-layer techniques for exploring brain networks at different temporal resolutions. In this section, we draw particular focus to the topic of multi-slice/layer community detection and its role in characterizing time-varying connectivity. Throughout both sections, we also comment on methodological limitations of these methods, the best practices for their application, and possible future directions. This review is written for the neuroimaging community, and so the literature we cover and the examples that we present are selected to be especially relevant for researchers working with MRI data (whether functional, diffusion, or structural). Nonetheless, our frank discussion of multi-scale methods and views are broadly relevant and applicable to researchers working with other data modalities (including EEG, MEG, ECOG, and fNIRS) and at other spatial scales in humans or other species.
	
	\section{Functional and structural brain networks}
	
	With MRI data, network nodes are almost always parcels of gray-matter voxels (sometimes the voxels, themselves, are used as nodes \cite{heuvel2008small}). Brain networks come in two basic flavors that differ from one another based on how connections are defined among nodes. \emph{Structural} or \emph{anatomical connectivity} (SC) networks refer to nodes linked by physical connections. With MRI data, these connections usually reflect white-matter fiber tracts reconstructed from the application of tractography algorithms to diffusion images. \emph{Functional connectivity} (FC) networks, on the other hand, refer to the strength of the statistical relationship between nodes' activity over time \cite{friston2011functional}. Usually this statistical relationship is operationalized as a Fisher-transformed correlation coefficient \cite{zalesky2012on} or a coherence measure \cite{zhang2016choosing}. Both SC and FC networks are represented with a connectivity matrix, $\mathbf{A}$, whose element $A_{ij}$ is equal to the connection weight between regions $i$ and $j$.
	
	\section{Multi-scale network analysis}
	
	Network analysis is the process of interrogating an SC or FC network using tools derived from graph theory in order to better understand its character. It is important to note that this type of analysis takes explicit account of the network architecture of SC and FC -- i.e. that the collective organization and configuration of connections gives rise to system-level behavior. It is therefore distinct from other techniques that examine SC and FC connection weights in isolation \cite{simpson2016disentangling}. Network science, which has existed as a field long before the advent of network neuroscience, has contributed a large number of measurements of a network that can help reveal its function, highlight influential nodes, and identify features that contribute to its robustness and vulnerability. The \emph{topological scale} at which a network is described depends upon what features of the network these measures highlight. Some measures are simple; a node's degree (or the weighted analog, strength) simply counts the number of connections incident on any node and can be interpreted as a measure of a node's influence, with high-degree nodes exhibiting the greatest influence \cite{takeuchi2015degree}. Degree is an example of a strictly local measure -- it characterizes only a single node. At the opposite end of the spectrum are measures that describe the organization of the network as a whole. A network's characteristic path length, for example, is the average number of steps it takes to go from one node to another. Short path lengths imply, at least in theory, that information can be quickly shared across the network \cite{santarnecchi2014efficiency,li2009brain}.
	
	Degree and path length, along with other local and global network measures, are useful for characterizing networks at their most extreme topological scales: at the level of a network's most commonly studied fundamental units (its nodes; although see \cite{giusti2016twos} and \cite{bassett2014cross} for alternatives) and the level of the network as a collective. Between these two scales lies a mesoscale, an intermediate scale at which a network can be characterized not in terms of local and global properties, but also in terms of differently sized clusters of nodes that adopt different types of configurations. It is at this mesoscale that we can observe community structure \cite{fortunato2010community}, cores and peripheries \cite{borgatti2000models}, and rich clubs \cite{colizza2006detecting}. It is essential to note that the mesoscale, unlike local and global scales, is defined as a \emph{range} of scales situated between two extremes. Therefore, mesoscale structures have the capacity to emerge, persist, and dissolve over multiple topological scales. In general, the detection of such structures is performed algorithmically, usually through the application of tools designed to detect specific types of mesoscale structure. As a simple illustration, consider a network with community structure. In the context of networks, communities refer to sub-networks (clusters of nodes and their edges) that are internally dense (many within-community edges) and externally sparse (few between-community edges) \cite{porter2009communities,newman2012communities}. One intuitive (and quite palatable) hypothesis is that brain networks are organized into hierarchical communities, meaning that communities at any particular scale can be sub-divided into smaller communities, which in turn can be further sub-divided, and so on \cite{meunier2010hierarchical,bassett2010efficient,hilgetag2014hierarchical}. This hierarchy can be ``cut'' at any particular level to obtain a single-scale description of the network's communities, but doing so ignores the richness engendered by the hierarchical nature of the communities. Similar arguments can be applied to other types of meso-scale organization, such as core-periphery \cite{bassett2013task} and rich clubs \cite{van2011rich}.
	
	In the following subsections, we review analysis techniques for the detection of mesoscale structure in brain networks, focusing on communities due to their inherent multi-scale nature. We pay particular attention to techniques that make it possible to detect community structure over a range of topological scales, thereby uncovering a richer, more detailed multi-scale description of brain networks.
	
	\begin{figure*}[t]
		\begin{center}
			\centerline{\includegraphics[width=1\textwidth]{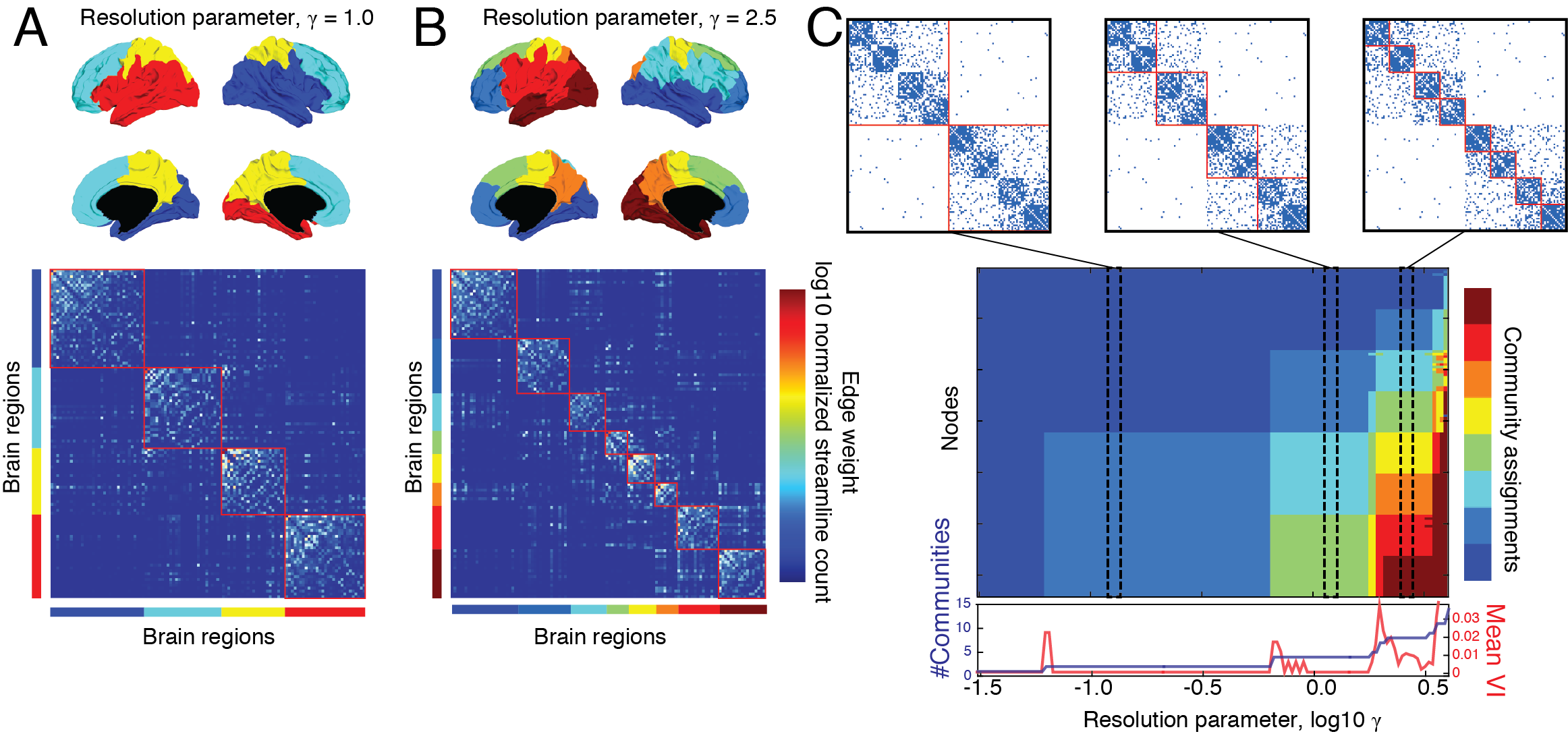}}
			\caption{\textbf{Schematic figure illustrating multi-scale community detection.} Networks exhibit community structure over a range of different topological scales. In panels \emph{(A)} and \emph{(B)} we show communities detected in a structural connectivity network at two different topological scales (the colors in the surface plots indicate the community to which each region is assigned). We investigate these scales by tuning the resolution parameter in modularity maximization (a common community detection approach) to $\gamma = 1$ and $\gamma = 2.5$. In panel \emph{(C)} we illustrate the multi-resolution approach for ``sweeping'' through a range of resolution parameters to detect communities at different scales, this time using a synthetic network constructed to have hierarchical community structure (hierarchical levels that divide the network into 2, 4, and 8 communities). To identify topological scales of interest (ranges of $\gamma$), we calculated the mean pairwise variation of information (VI) of all partitions detected at each value of $\gamma$. Low values of VI indicate that on average the detected partitions were similar to one another. The metric VI achieves local minima at scales that uncover the planted hierarchical communities; at values of $\gamma$ where none of the planted hierarchical communities are detected, VI takes on non-zero values, indicating lack of consensus across detected partitions and highlighting values of $\gamma$ at which community structure is not present.}
			\label{figure:multiScaleCommunityDetection}
		\end{center}
	\end{figure*}
	
	\subsection{Multi-scale community structure}
	
	Local and global properties of networks are straightforward to compute because the units of analysis -- individual nodes and the whole network -- are immediately evident and require no additional search. Mesoscale structure, however, is not always evident. Its presence or absence in a network depends on the configuration of edges among the network's nodes -- that is, the network's \emph{topology}. Real-world networks are composed of many nodes and edges arranged in complex patterns that can obscure structural regularities. Due to this complexity, if one wishes to observe mesoscale structure in networks, one must search for it algorithmically. In the case of community structure \cite{meunier2010hierarchical,sporns2016modular}, there is no shortage of algorithms for doing so. They range both in terms of how they define communities and also their computational complexity \cite{palla2005uncovering, ahn2010link, rosvall2008maps, delvenne2010stability, karrer2011stochastic}. Whether the plurality of methods is viewed as a shortcoming or an advantage, the enterprise of community detection is one of the better-developed and continually-growing sub-fields of network analysis \cite{fortunato2010community,fortunato2016community}.
	
	While each community detection technique offers its own unique perspective on how to identify communities in networks, the method that is most widely used and arguably the most versatile is \emph{modularity maximization} \cite{newman2004finding}. Modularity maximization partitions a network's nodes into communities so as to maximize an objective function known as the \emph{modularity} (or just ``$Q$''). The modularity function compares the observed pattern of connections in a network against the pattern that would be expected under a specified null model of network connectivity. That is, the weight of each existing edge is directly compared against the weight of the same edge if connections were formed under the null model. Some of the observed connections will be unlikely to exist under the null model or will be stronger than the null model would predict. Modularity maximization tries to place as many of the stronger-than-expected connections within communities as possible.
	
	More formally, if the weight of the observed and expected connection between nodes $i$ and $j$ are given by $A_{ij}$ and $P_{ij}$, respectively, and $\sigma_i \in [1, \ldots, K]$ indicates to which of $K$ communities node $i$ is assigned, then the modularity can be calculated as:
	
	\begin{equation}
	Q=\sum_{ij} [A_{ij} - P_{ij}] \delta (\sigma_i \sigma_j), 
	\end{equation}
	
	\noindent where $\delta(\cdot \cdot)$ is the Kronecker delta function and is equal to $1$ if its arguments are the same and $0$ otherwise. Multiple methods exist to actually maximize $Q$, but in the end they all result in an estimate of a network's community structure: a \emph{partition} of the network nodes into communities.
	
	The number and size of communities in the partition with the biggest $Q$ represent the communities present in the network, right? Unfortunately, the answer to this question is ``not always.'' Modularity and other similar quality functions exhibit a ``resolution limit'' that limits the size of detectable communities \cite{fortunato2007resolution}; communities smaller than some size, even if they otherwise adhere to our intuition of a community, are mathematically undetectable. In order to detect communities of all sizes, modularity has been extended in recent years to include a resolution parameter, $\gamma$, that can be tuned to uncover communities of different size \cite{reichardt2006statistical}. The augmented modularity equation then reads:
	
	\begin{equation}
	Q(\gamma)=\sum_{ij} [A_{ij} - \gamma P_{ij}] \delta (\sigma_i \sigma_j) .
	\end{equation}
	
	\noindent The resolution parameter was initially introduced as a technique for circumventing the resolution limit. Inadvertently, it has contributed to the versatility of the modularity measure. The resolution parameter effectively acts as a tuning knob, making it possible to obtain estimates of small communities when it is at one setting and larger communities when it is at another setting: when $\gamma$ is big or small maximizing modularity will return correspondingly small or large communities. If we smoothly tune the resolution parameter from one extreme to the other, we can effectively obtain estimates of a network's community structure, all the way from the coarsest scale at which all network nodes fall into the same community up through the finest scale where network nodes form singleton communities. Varying the resolution parameter to highlight communities of different sizes is known as \emph{multi-scale community detection} \cite{fenn2009dynamic}. It should be noted that there exist possible definitions of modularity functions that do not suffer from resolution limits in the first place \cite{traag2011narrow}. A full discussion of these functions is beyond the scope of this review.
	
	\subsubsection{Multi-scale community structure in the neuroimaging literature}
	
	Multi-scale analyses of real-world networks have revealed known structural motifs in proteins \cite{delmotte2011protein,delvenne2010stability}, dynamic patterns in financial systems \cite{fenn2009dynamic,fenn2012dynamical}, and ``force chains'' in physical systems of particles \cite{bassett2015extraction}. Most studies of community structure in brain networks, however, have focused on communities at a single scale \cite{hagmann2008mapping,power2011functional,gu2015emergence} or, in the event that investigators wish to examine multiple scales, have resorted to heuristics such as recursive partitioning \cite{he2009uncovering,bassett2010efficient}, edge thresholding \cite{power2011functional}, or by accepting sub-optimal solutions through the modification of existing algorithms \cite{meunier2010modular}. The multi-scale modularity maximization approach and related techniques \cite{delvenne2010stability, schaub2012markov, kheirkhahzadeh2016efficient} can seamlessly scan all topological scales by tuning the resolution parameter, which entails no additional assumptions. While single-scale approaches to community detection are not fundamentally wrong, they miss out on the richness that may be present at other scales. For example, a single-scale estimate of the community structure for a hierarchically modular network would detect only one of the hierarchical scales present in the system.
	
	Nonetheless, there is a growing number of studies that have employed multi-scale community detection techniques \cite{rubinov2015wiring}. Some of these studies used the multi-scale approach to identify single-scale modules, but at a resolution parameter that differs from the default ($\gamma = 1$) \cite{gu2015controllability, betzel2016modular, nicolini2016modular}. In other words, they obtained estimates of community structure over multiple scales and defined a secondary objective function that, when optimized, identified from among that set of partitions a scale at which to focus on. Other approaches have explicitly set out to compare community structure detected at different resolutions. In the aging literature, for example, a number of studies have reported that communities become less segregated across the human lifespan \cite{chan2014decreased,betzel2014changes}. In a recent study, however, the authors analyzed the community structure of resting-state FC networks across the lifespan and at different values of $\gamma$ \cite{betzel2015functional}. They showed that community structure, and specifically the extent to which communities are segregated from one another, exhibits an interaction between age and scale; smaller communities become less segregated with age, while larger communities become increasingly segregated. However, had the authors only explored community structure at a single topological scale, they would have never observed the reported interaction.
	
	Other studies have estimated multi-scale community structure towards more theoretical ends. For example, in \cite{lohse2014resolving}, the authors characterize different spatial and topological properties of anatomical brain networks as a function of $\gamma$, and use a measure of community radius \cite{doron2012dynamic} to show that large communities (as measured by the number of nodes) are embedded in large physical spaces. This mapping of a large topological entity to a large physical entity is not required of networked systems \cite{barthelemy2011spatial}, and its existence suggests the presence of non-trivial constraints on the embedding of the brain's network architecture within the confines of the human skull \cite{bullmore2012economy}. Indeed, the multiscale nature of the brain's modular architecture is strikingly similar to the hierarchical modularity observed in large-scale integrated circuits, whose abstract (and rather complex) topology has been mapped cost-efficiently (meaning with a predominance of short wires) into the two-dimensional space of a computer chip \cite{bassett2010efficient,klimm2014resolving}. This efficient mapping can be uncovered by testing for the presence of Rentian scaling \cite{sperry2016rentian}, a property by which the number of edges crossing the boundary of a spatial parcel of the network scales logarithmically with the number of nodes inside the parcel. Hierarchically modular networks -- including the human brain, the \emph{C. elegans} neuronal network, and even the London underground -- that have been efficiently embedded into physical space commonly display Rentian scaling, while those that have not been efficiently embedded do not show this property.

	\subsubsection{Implementation and practical considerations}
	
	Community detection, generally, is easy to do but difficult to do well \cite{fortunato2016community}. Modularity maximization for community detection begins with the assumption that the network is modular \cite{lancichinetti2011finding}, and as a technique is prone to false positives \cite{guimera2004modularity}. Moreover, detecting the globally optimal partition is computationally intractable \cite{good2010performance}, the most popular algorithm for maximizing modularity generates variable output \cite{blondel2008fast}, and the composition of detected communities can be biased by the overall density of the network \cite{fortunato2007resolution}. These are issues associated with modularity maximization \emph{before} sweeping $\gamma$. Adding the resolution parameter can further amplify these complications; these issues are manifest at \emph{every} level of $\gamma$. How can the prospect of multi-scale modularity maximization be performed in a principled, careful, and thoughtful way?
	
	\subsubsection*{Selecting the resolution parameter}
	One of the most important issues is to select the topological scale(s) of interest, which is tantamount to focusing on a subset of $\gamma$ values. Without prior knowledge of the number and size of communities, there is no good rationale for preferring one value of $\gamma$ over another (including $\gamma = 1$). There are, however, a few approaches described in the existing literature for selecting a scale of interest from among the communities detected over a range of $\gamma$ values. Intuitively, if a network's organization at a particular scale is truly well-described by communities, then we might also believe that our algorithms will easily detect this organization. In this case, the known variability in the output of some modularity maximization techniques \cite{blondel2008fast} can actually work in our favor. When variability is low -- i.e. the algorithm converges to similar community structure estimates over multiple runs -- it might be indicative of especially well-defined communities. Under this assumption, we repeatedly maximize modularity at different values of $\gamma$ and calculate the pairwise similarity of the detected communities \cite{doron2012dynamic}. We can then focus on community structure detected at $\gamma$ values where the similarity is great (and variability low) (See \cite{betzel2016modular,gu2015controllability,chai2016functional,mattar2015functional} for examples where this approach has been applied). Similarity of partitions can be estimated using a number of measures such as normalized mutual information \cite{lancichinetti2009detecting}, variation of information \cite{meilua2003comparing}, or the $z$-score of the Rand coefficient \cite{traud2011comparing}.
	
	Other approaches have also been suggested. One possibility is to use statistical arguments to focus on specific scales of $\gamma$. For example, we could estimate the probability of observing a community of a particular size by chance, and then focus on the scale where the detected communities' sizes deviate most from chance \cite{traag2013significant}. Another possibility assumes that ``good'' community structure is not fleeting -- i.e. that it should persist over some range of $\gamma$ \cite{fenn2009dynamic}. Under this assumption we can calculate the average similarity between partitions detected at every pair of $\gamma$ values and cluster the resulting similarity (or distance) matrix. The clusters correspond to collections of detected partitions that are all highly similar to one another -- the absence of clusters suggests that if community structure exists at different scales, then it is short-lived and possibly of less interest \cite{lambiotte2014random}. At the very least, in the event that one does not wish to scan multiple topological scales, a good method for demonstrating the robustness of a result that depends upon the composition of detected communities is to vary $\gamma$ slightly from the selected value to verify that community structure is consistent (see, for example: \cite{betzel2016positive}).

	\subsubsection*{Consensus community structure and communities of interest}
	Choosing the $\gamma$ value(s) at which to analyze a network's community structure is the first hurdle. There remain the unresolved questions of how to define \emph{consensus communities} that are representative over a group of partitions and how to determine whether all (or just some) of the detected consensus communities are of interest (the group of partitions could come from multiple optimizations of a modularity maximization algorithm or a collection of partitions obtained from many individuals). There are now multiple approaches for choosing a consensus partition, including ``similarity maximization'' (choosing the consensus partition as the one with greatest average similarity to the other partitions) \cite{doron2012dynamic} and variants of the ``association-recluster'' framework (using a clustering algorithm to find consensus communities in a co-occurence or association matrix that stores the frequency with which nodes co-occur in a community over an ensemble of partitions) \cite{lancichinetti2012consensus,bassett2013robust,betzel2016modular}. Because these approaches are now well-known and widely-used, we will not discuss them further here.
	
	We do, however, find it prudent to discuss the final question: ``should we analyze all the communities in the partition?'' The notion of defining a partition in which all nodes get assigned to one community or another presupposes that this type of structure exists in the first place. Is this a reasonable assumption? The presence of hubs \cite{hagmann2008mapping} and rich-clubs \cite{van2011rich} suggests that at least some brain network nodes fail to strictly adhere to the community template -- hub nodes, by definition, are highly connected and span multiple modules. In short, maximizing modularity \emph{always} partitions the network into clusters, but are all the clusters really communities? There are multiple ways to address this question. One possibility is, again, to invoke a statistical argument and ignore communities with properties consistent with what you might expect by chance. For instance, you could calculate the modularity contribution made by each community (defined in \cite{bassett2012influence} and applied in \cite{betzel2014changes,betzel2016modular}) and compare the observed values against a random null model (e.g., permute the community labels and recalculate modularity contributions, optimize modularity for rewired networks and compare the observed modularity to that of the randomized networks). The gold standard technique, however, would be a tool that does not force all nodes to be in a community and only detects communities that are inconsistent with a random null model. Such a tool exists in the form of the OSLOM algorithm \cite{lancichinetti2011finding}, which works by first identifying the worst node in a community (i.e. the one with the fewest within-community connections). Next, the community is assigned a ``$C$-score'' defined as the probability of observing a node in the same community that makes more within-community connections than expected in a random network. To the best of author's knowledge and at the time of writing this review, OSLOM has not yet been applied to brain network data.
	
	In this section, we highlighted the fact that networks can exhibit non-random organization across a range of topological scales, from that of individual nodes up to the entire network. To develop a more complete understanding of the network's organization and develop deeper insight into its function, we argue that it is essential to focus not only on one single scale, but to embrace the multi-scale topological nature of brain networks and characterize brain networks using appropriately multi-scale tools. The result is a richer picture of a brain network. That added richness may be necessary to form a deeper understanding of how brain network structure is associated with human behavior and cognition, and ultimately how it is altered in disease.

	\subsection{Multi-scale rich club and core-periphery organization}
	
	\begin{figure*}[t]
		\begin{center}
			\centerline{\includegraphics[width=0.54\textwidth]{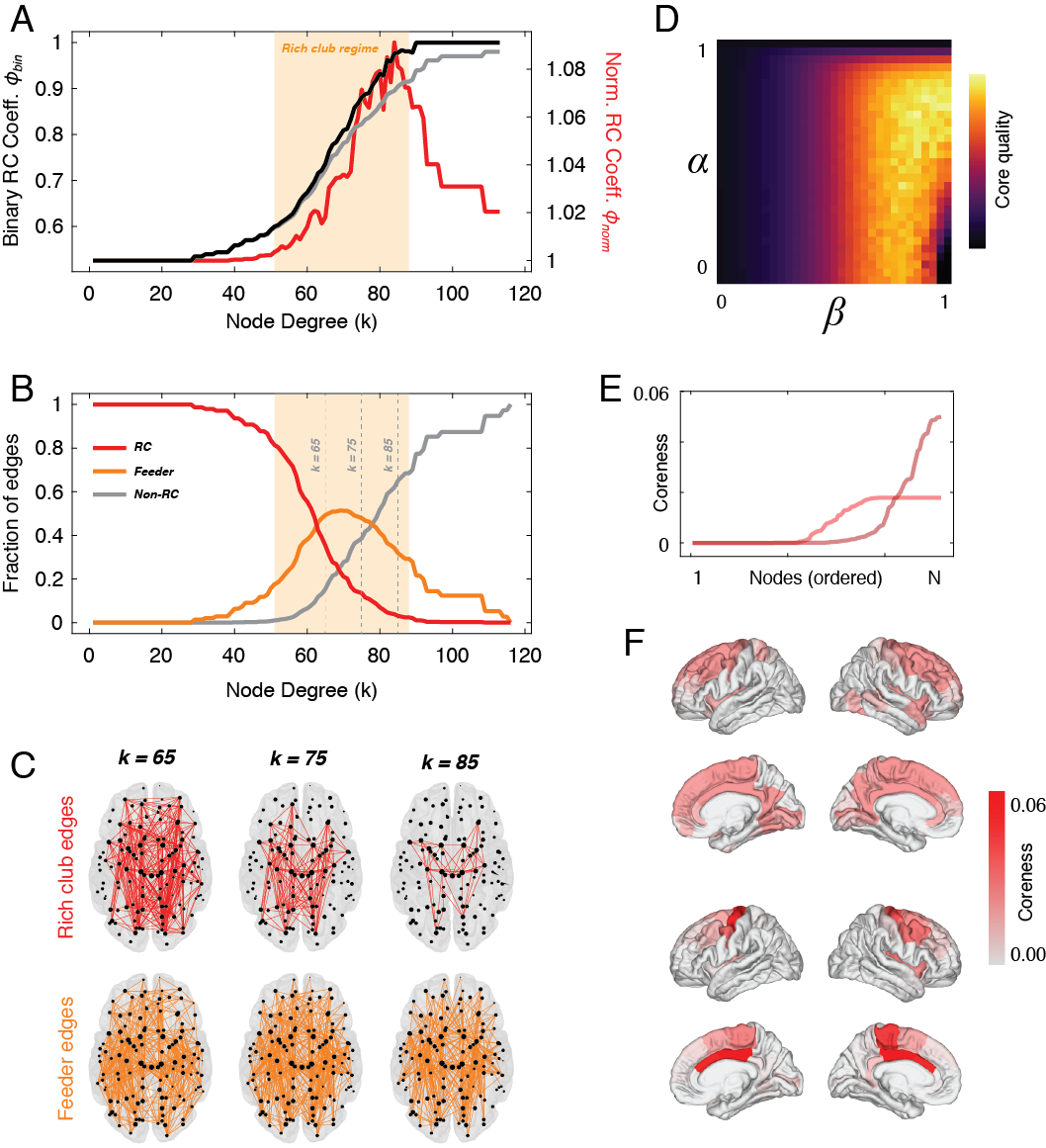}}
			\caption{\textbf{Multi-scale rich club and core-periphery analysis}. (\emph{A}) The rich club coefficient, $\phi_{bin}$, for the observed network (black) and the mean over an ensemble of random networks (gray) as a function of node degree, $k$. The ratio of these two measures defines the normalized rich club coefficient, $\phi_{norm}$. Values of $k$ for which the observed rich club coefficient is statistically greater than that of a random network define the rich club regime. (\emph{B}) Most studies focus on a rich club defined at a single $k$ value and use it to classify edges as ``rich club'' (rich node to rich node), ``feeder'' (rich node to non-rich node), or ``non-rich club'' (non-rich node to non-rich node). The number of edges assigned to each class is highly dependent upon the $k$ at which the rich club is defined. (\emph{C}) We show edge classifications at three different values of $k$, in order to highlight that classifications (and the subsequent interpretation) can vary dramatically, even across statistically significant rich clubs. (\emph{D}) Core-periphery classification can be performed using a parameterized model \cite{rombach2014core}. The parameters $(\alpha,\beta)$ determine the size of the core relative to the periphery and how sharply the two are divided from one another \cite{bassett2013task}. At different parameter values the model identifies different cores and different peripheries, and assigns each node a ``coreness'' score. (\emph{E}) As an example, we show two sets of coreness scores ordered from smallest to largest. The two sets vary in terms of the core size and constitution. (\emph{F}) For the same two sets, we show the topographic distribution of coreness scores. Note: In both the rich club and core-periphery examples, the network studied was a structural network used in a previous study \cite{betzel2016optimally}.} \label{richClubMultiScale}
		\end{center}
	\end{figure*}
	
	In addition to community structure, networks can exhibit a range of mesoscale organizations. These include rich club and core-periphery structure, both of which have been investigated in the context of brain networks. While not the explicit focus of this review, we felt that we would be remiss not to briefly mention the available tools to study multi-scale rich club and core-periphery organization. 
	
	We recall that a rich club is a group of hubs (high degree, high strength nodes) that are also densely interconnected to one another \cite{colizza2006detecting,opsahl2008prominence}. Rich clubs are hypothesized to act as integrative structures in SC networks by linking modules to one another and facilitating rapid transmission of information \cite{van2011rich}. Core-periphery structure is a related concept, which assumes that the network consists of one (or a few) dense cores, with which peripheral nodes interact, though the peripheral nodes rarely interact with one another \cite{borgatti2000models,holme2005core,rombach2014core}. Similar to rich clubs, cores play an integrative role, serving as a locus for different brain regions to link up and exchange information.
	
	Similar to communities, there is a tendency in the network science literature to concoct binary assignments of nodes as either belonging to or not belonging to a network's cores and rich clubs. This dichotomy aids in the interpretation of results, but ultimately belies the complexity and richness of core-periphery and rich club organization in a network, both of which can persist over multiple topological scales. Whereas communities can be identified by maximizing a modularity function, rich clubs are detected by calculating a rich-club coefficient, $\phi (k)$, which measures the density of connections among nodes with degree $k$ or greater (Fig.~\ref{richClubMultiScale}A). If this coefficient is greater than what would be expected under a random network model, there is evidence that the rich club is statistically significant. In practice, there is nearly always a plurality of statistically significant rich clubs, and hence a plurality of rich club nodes. The absence of a singular rich club gives rise to multiple complementary views of how hub nodes interact with one another and how they contribute to brain function (Fig.~\ref{richClubMultiScale}B,C). A similar argument applies for core-periphery structure, where nodes can be more or less core- or periphery-like in a graded sense, defying the dichotomy of being one or the other (Fig.~\ref{richClubMultiScale}D-F). 
	
	Is there a practical way to assess these types of multi-scale rich clubs and core-periphery structures? In the case of rich clubs, one natural solution is to report the range of statistically significant rich clubs and characterize the composition of rich clubs across that range. In the case of core-periphery organization, one can study a parameterized landscape of core-periphery architecture, offering a continuous description of cores of different sizes, and with differing softness of the boundary between the core-like nodes and the periphery-like nodes \cite{rombach2014core,bassett2013task} (Fig.~\ref{richClubMultiScale}D-F). These and other approaches that are similar in spirit may offer additional insights into the multi-scale architecture of the brain in a manner that complements the assessment of heirarchical community structure described in detail in earlier sections.
	
	\subsection{Multi-scale temporal networks}
	
	At this point, we take a step back and note that brain networks, both functional and structural, are not static but instead fluctuate over timescales ranging from the sub-second \cite{kopell2014beyond,calhoun2014chronnectome} to the lifespan \cite{di2014unraveling}. These fluctuations in network organization, especially over short timescales ($<$ that of a single scan session), have become frequent topics of investigation \cite{bassett2011dynamic,allen2012tracking, bassett2015learning, zalesky2014time, betzel2016dynamic}.
	
	How do we study a network that changes over multiple timescales? One promising approach is to use multi-layer network models of temporal networks \cite{kivela2014multilayer, de2013mathematical}. The multi-layer network model is flexible enough to deal with networks that vary along dimensions other than time \cite{muldoon2016network}, but when applied to temporal networks it treats estimates of the network's topology at different time points as ``layers''. For example, a layer could represent a functional network estimated from a few minutes of observational data acquired during an fMRI BOLD scan \cite{telesford2016detection} or it could represent the structural connectivity of an individual participant at a particular age in a developmental or lifespan study \cite{betzel2015functional}. Whereas traditional network analysis would characterize each layer independently of one another, multi-layer network analysis treats the collection of layers as a single object, characterizing its structure as a whole to explicitly bridge multiple temporal scales. Equally important, the multi-layer network model is agnostic (from a mathematical perspective) to the timescales represented by the layers, and can therefore accommodate virtually any timescale made accessible using neuroimaging technologies.
	
	\begin{figure*}[t]
		\begin{center}
			\centerline{\includegraphics[width=1\textwidth]{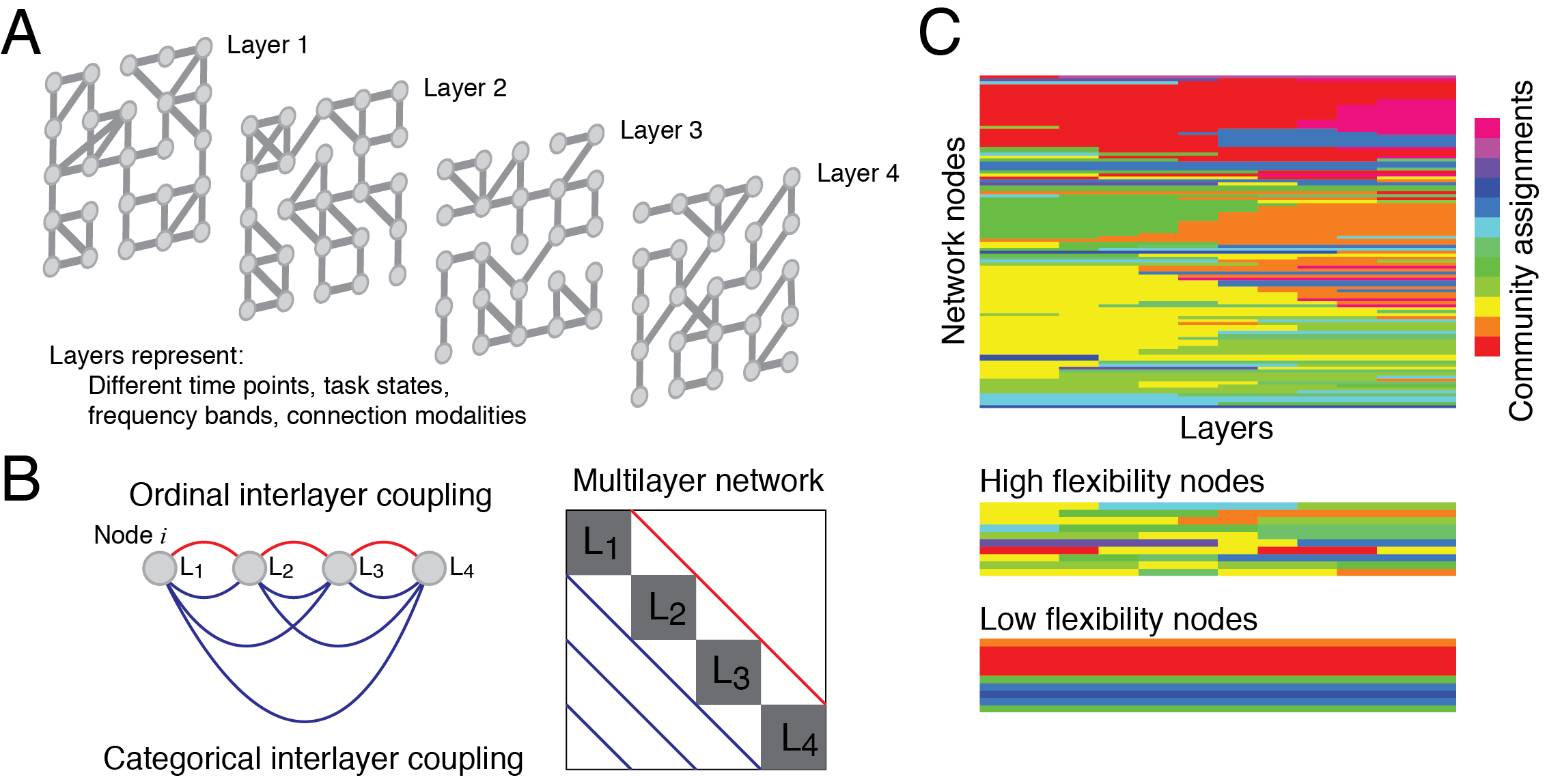}}
			\caption{\textbf{Schematic figure illustrating multi-layer network construction and community detection.} Individual networks can be combined in a meaningful way to form a multi-layer network. In panel \emph{(A)} we show four example networks, each of which contains the same 25 nodes but arranged in different configurations. The links in these networks could represent fluctuating functional connections over time (e.g., within a single scan or over development), connections estimated during different tasks, different frequency bands, or different connection modalities (e.g., structural connections weighted by streamline count or fractional anisotropy or functional connections measured as correlations or coherence). \emph{(B)} To combine individual layers, links are added from node $i$ to itself across layers. These links can be added ordinally, linking a node to itself in adjacent layers, or categorically, linking a node to itself across all layers. The result is a multi-layer network. \emph{(C)} Multi-layer networks can be analyzed using many now-standard measures in network science, including -- but not limited to -- community detection algorithms. The resulting estimate of communities allows us to track the formation and dissolution of communities across layers and report properties of individual nodes -- e.g., their flexiblity, which measures how frequently a node changes its community assignment.}
			\label{figure:multiLayerNetworks}
		\end{center}
	\end{figure*}
	
	\subsubsection{Multi-scale, multi-layer network analysis}
	
	Most of the familiar network measurements have been generalized so that they can be computed on a multi-layer network. For example, path length, clustering, and some centrality measures are all easily calculated \cite{kivela2014multilayer}. While a few recent studies have begun to investigate these measures in multi-frequency brain networks \cite{de2016mapping, brookes2016multi, battiston2016multilayer}, the most widely used multi-layer measure in network neuroscience is that of multi-layer, multi-scale community detection \cite{mucha2010community}. Though there are several different approaches for detecting communities in temporal networks, including non-negative matrix factorization \cite{gauvin2014detecting, ponce2015resting}, and hidden Markov models \cite{robinson2015dynamic}, the most popular is multi-layer modularity maximization, which represents a powerful extension of the standard modularity maximization framework that makes it possible to uncover communities across layers (i.e., time, in the case of temporal networks). The multi-layer analog resolves several important issues. First, it confers further flexibility to the multi-layer network model by making accessible familiar methods.  Communities can, of course, be calculated for each layer independently. This unfortunately gives rise to ambiguities regarding the continuation of communities from one layer to the next. The second advantage of the multi-layer model is that by estimating the community structure of all layers simultaneously such ambiguities are effectively resolved. Third, it opens the possibility of defining new measures for characterizing the flow of communities across layers \cite{bassett2013robust,mattar2015functional,papadopoulos2016evolution}. For example, the measure ``flexibility'' quantifies how frequently a brain region changes its community assignment from one layer to the next \cite{bassett2011dynamic}. Increased flexibility has been associated with learning \cite{bassett2011dynamic}, increased executive function \cite{braun2015dynamic}, aging \cite{betzel2015functional}, and positive mood, novelty of experience, and fatigue \cite{betzel2016positive}. Additionally, it can also be used to reveal a temporally stable core of primary sensory systems along with a flexible periphery of higher-order cognitive systems \cite{bassett2013task} offering an architecture thought to be particularly conducive to flexible cognitive control \cite{fedorenko2014reworking}. Other statistics including ``promiscuity'' offer distinct quantifications of meso-scale network reconfiguration \cite{papadopoulos2016evolution}.
	
	Importantly, multi-layer modularity maximization includes a resolution parameter, $\gamma$, that functions in an analogous manner to the resolution parameter in single-layer community detection. In conjunction with the multi-layer framework, which facilitates the investigation of temporal networks, the resolution parameter gives a researcher the option of incorporating multiple topological scales into a temporal analysis of networks.
	
	\subsubsection{Practical considerations}
	
	The multi-layer model can accommodate many different types of data collected over multiple timescales. This freedom comes at a cost, however. In order to consider all layers as forming a single multi-layer network object, it is currently a necessity to, either manually or in some data-driven way, add artificial links between layers. Broadly, there are two strategies for this approach. The first assumes that layers are not ordinally related to one another -- i.e. layers have no temporal precedence with respect to one another; a permutation of the order of layers results in effectively the same network. If these assumptions hold (e.g., if layers represent connectivity matrices obtained from different task states), then it makes sense to \emph{categorically} link layers to one another \cite{cole2014intrinsic,mattar2015functional}. If the layers exhibit an \emph{ordinal} relationship, then it makes more sense to link node $i$ in layer $s$ to its temporally adjacent layers, $s - 1$ and $s + 1$ \cite{chai2016functional}. The decision to choose one approach over the other can, of course, influence whatever measurement is being made on the network. Currently, it is standard practice (at least in network neuroscience) to add ordinal links when dealing with temporal networks.
	
	Even with sound rationale for selecting one linking procedure over the other, there still remains the difficult decision of how to assign the inter-layer links a weight. Again, how these weights are selected can have an effect on whatever measure is being computed. Without strong evidence to select one weighting scheme over another, interlayer links are usually assigned the same value, $\omega$, that is sometimes varied over a narrow range. Ideally, there would be a principled, data-driven approach for selecting this value.
	
	\subsection{Multi-scale spatial networks}
	
	The explosion of network science into different scientific communities can be attributed, in part, to the fact that it provides a set of tools that can be applied to network data of all types. In this review, we focused on brain networks derived from functional and diffusion MRI, the modalities most often used in the neuroimaging community. The networks constructed from these data span spatial scales ranging from that of individual voxels up to that of the whole brain. The nature of MRI, however, makes it virtually impossible to construct brain networks at finer scales, such as the level of individual cells or that of neuronal populations. Other spatial scales are, of course, accessible using alternative imaging modalities. For example, optical imaging has delineated cellular-level networks of mouse retina \cite{helmstaedter2013connectomic, lee2016anatomy} as well as of model organisms like the nematode, \emph{C. elegans} \cite{jarrell2012connectome}, or \emph{drosophila} \cite{takemura2013visual}. Large-scale tract-tracing and fluorescent labeling techniques have proven useful in uncovering networks at an intermediate scale -- detecting axonal projections between local processing units in \emph{drosophila} \cite{shih2015connectomics}, and brain areas in mouse \cite{oh2014mesoscale} and macaque \cite{markov2012weighted}. Additionally, meta-analytic studies that aggregate and summarize the results of individual tract-tracing experiments have produced convergent maps of macaque \cite{stephan2001advanced} and rat \cite{bota2015architecture} network architecture. At these scales, the details of what each node and edge represent differ from that of whole-brain human networks. Nonetheless, the same network analysis tools can be brought to bear on these networks to reveal their organization and gain insight into their function. As microscale imaging tools become more common, and existing tools more refined, capable of handling higher throughput, and imaging greater volumes, they will be able to offer novel insights into how the multi-scale spatial network structure of the brain relates to cognition and behavior. An important step in advancing the field of network neuroscience is understanding, specifically, how network properties at one spatial scale are related to properties at another \cite{van2015bridging}.
	
	Presently, of course, the analysis of human brain networks is limited by the spatial granularity of the individual voxel. Even with this lower bound on the size of brain network nodes, it is possible to probe multiple spatial scales using MRI data. The most obvious manner in which spatial scale can be examined is in the choice of brain parcellation. MRI acquisitions return observations at the level of individual voxels. Voxels may be noisy, suffer from signal dropout, and due to their large number may present computational challenges to conduct analyses at that scale. For these reasons, it has become common to aggregate voxels into parcels or regions of interest; rather than focus on any particular voxel, this allows us to focus on the average properties of parcels \cite{de2013parcellation}.
	
	The number of alternative parcellations is ever-growing, with each new parcellation presenting a new criteria -- e.g., spatial variation in functional connectivity, myelination, cytoarchitectonics, etc. -- for grouping voxels together into regions \cite{destrieux2010automatic, yeo2011organization, gordon2014generation, glasser2015multi,hermundstad2013structural}. The number of parcels ranges from $\approx$1000 \cite{cammoun2012mapping, diez2015novel} to around 60 for the whole brain, representing a massive reduction from the tens of thousands of voxels typically imaged. Looking at parcellations of the brain from the voxel-level down to the coarsest set of parcels, we can examine different spatial scales of the brain. One of the findings that has come from a detailed comparison of spatial scales is that the choice of parcellation will tend to have implications for the network's topology \cite{wang2009parcellation, zalesky2010whole}. For this reason, it is advised to verify that any particular result is not driven by the particular choice of parcellation; it should be reproducible (at least qualitatively) using a different set of parcels \cite{bassett2011conserved}. A potentially interesting avenue for future work in this area is to apply multi-scale community detection to voxel-level networks to generate parcellations of the brain at different resolutions \cite{bellec2010multi}. The parcellation-based approach for studying different spatial scales can be used to investigate and sub-divide specific brain areas, rather than the entire brain \cite{rosenthal2016stimulus}. For example, in one recent study, owed to the retinotopic organization of the visual cortices, the authors were able to identify distinct parcels based on their connectivity patterns \cite{dawson2016partial}.
	
	\section{Conclusion and future directions}
	
	This review deals with the topic of multi-scale brain networks. We discuss tools for performing multi-scale network analysis, their application to time-resolved networks that highlight network-level fluctuations across multiple temporal resolutions, and finally touch briefly on how different spatial scales of analysis are making an impact on the field of network neuroscience. The results of network analyses at different scales can be seen as both redundant and complementary. In some sense, we expect to find similar network properties across scales \cite{van2016comparative} -- the same energetic and spatial constraints that shape network structure at the scale of brain regions and areas are at play at the cellular-level \cite{betzel2016generative,henriksen2016simple, vertes2012simple}. On the other hand, the function of network nodes and circuits as well as their biophysical attributes likely depend critically upon the scale at which a network is constructed and analyzed. Accordingly, we might also expect networks to be optimized to perform scale-specific functions \cite{marblestone2016towards}, and studying a particular scale gives us a unique insight into the network architecture underpinning those functions. Ultimately, network neuroscience will need both approaches -- an understanding of network function and organization at specific scales, as well as a map that bridges multiple different spatial, temporal, and topological scales.

	\section{Acknowledgements}
	
	We thank Arian Ashourvan and Lia Papadopoulos for helpful feedback on an earlier verson of this manuscript. DSB would also like to acknowledge support  from the John D. and Catherine T. MacArthur Foundation, the Alfred P. Sloan Foundation, the  Army Research Laboratory and the Army Research Office through contract numbers W911NF-10-2-0022 and W911NF-14-1-0679, the National Institute of Health (2-R01-DC-009209-11,1R01HD086888-01, R01-MH107235, R01-MH107703, and R21-M MH-106799), the Office of Naval Research, and the National Science Foundation (BCS-1441502, BCS-1430087, PHY-1554488, and BCS-1631550). The content is solely the responsibility of the authors and does not necessarily represent the official views of any of the funding agencies. 
	
	Depiction of gray-matter development in \ref{figure:overviewFigure} was reproduced from \cite{gogtay2004dynamic}. Copyright (2004) National Academy of Sciences, U.S.A.

	\newpage
	\section*{References}
	\bibliography{ni_multiscale}

\end{document}